# Latent Semantic Learning with Structured Sparse Representation for Human Action Recognition

Zhiwu Lu and Yuxin Peng*

*Abstract*—This paper proposes a novel latent semantic learning method for extracting high-level features (i.e. latent semantics) from a large vocabulary of abundant mid-level features (i.e. visual keywords) with structured sparse representation, which can help to bridge the semantic gap in the challenging task of human action recognition. To discover the manifold structure of mid-level features, we develop a spectral embedding approach to latent semantic learning based on $L_1$-graph, without the need to tune any parameter for graph construction as a key step of manifold learning. More importantly, we construct the $L_1$-graph with structured sparse representation, which can be obtained by structured sparse coding with its structured sparsity ensured by novel $L_1$-norm hypergraph regularization over mid-level features. In the new embedding space, we learn latent semantics automatically from abundant mid-level features through spectral clustering. The learnt latent semantics can be readily used for human action recognition with SVM by defining a histogram intersection kernel. Different from the traditional latent semantic analysis based on topic models, our latent semantic learning method can explore the manifold structure of mid-level features in both $L_1$-graph construction and spectral embedding, which results in compact but discriminative high-level features. The experimental results on the commonly used KTH action dataset and unconstrained YouTube action dataset show the superior performance of our method.

## I. INTRODUCTION

Automatic recognition of human actions in videos has a wide range of applications such as video summarization, human-computer interaction, and activity surveillance. Although many impressive results have been reported on human action recognition, it still remains a challenging problem [1] owing to viewpoint changes, occlusions, and background clutters. In the literature, one direct strategy is to measure how humans are moving in the scene, using the techniques for tracking or body pose estimation [2]–[4]. However, a distinct limitation of this strategy is that it requires reliable tracking or body pose estimation, which is difficult for realistic videos. Another more effective strategy adopts an intermediate representation based on spatio-temporal interest points [5]–[7] to bridge the semantic gap between low-level spatio-temporal features and high-level action categories. In particular, recent work has shown promising results when the local spatio-temporal descriptors are used for bag-of-words (BOW) models [8]–[11], where the local features are quantized to form a visual vocabulary and each video clip is thus summarized as a histogram of visual keywords. In the following, we refer to the visual keywords as mid-level features to distinguish them from the low-level features and high-level action categories.

However, this BOW representation may suffer from the redundancy of mid-level features, since typically thousands of visual keywords are formed to obtain better performance on a relatively large action dataset [12]. Here, it should be noted that the large vocabulary size means that the BOW representation would incur large time cost in not only vocabulary formation but also later action recognition. Moreover, the mid-level features are applied to human action recognition independently and mainly the first-order statistics is considered. Intuitively, the higher-order semantic correlation between mid-level features is very useful for bridging the semantic gap in human action recognition. Although the semantic information can be incorporated into the visual vocabulary using either local descriptor annotation or video annotation, the manual labeling is too expensive and tedious for a large action dataset. Therefore, to reduce the redundancy of mid-level features, this paper focuses on automatically extracting high-level features (or latent semantics) that are compact in size but more discriminative in terms of descriptive power.

Previously, several unsupervised methods [13], [14] have been developed to learn latent semantics based on topic models, such as probabilistic latent semantic analysis (PLSA) [15] and latent Dirichlet allocation (LDA) [16]. A mixture of latent topics is used to model each video, and the topics are learnt as multinomial distributions of mid-level features. Moreover, information theory has also been applied to latent semantic analysis for human action recognition in [17], [18]. The success of these models may be due to the fact that the semantically similar mid-level features generally have a higher probability of co-occurring in a video across the entire dataset. It should be noted that, besides this simple co-occurring information, there also exists more complicated semantically similar information, e.g., the mid-level features generated from similar video contents tend to lie in the same geometric or manifold structure. However, this intrinsic information is not considered by the latent topic or information theoretic models [13], [14], [17], [18]. In the literature, very few attempts have been made to explicitly preserve the manifold geometry of the mid-level feature space when learning high-

The authors are with the Institute of Computer Science and Technology, Peking University, Beijing 100871, China (e-mail: luzhiwu@icst.pku.edu.cn, pengyuxin@icst.pku.edu.cn).
* Corresponding author.

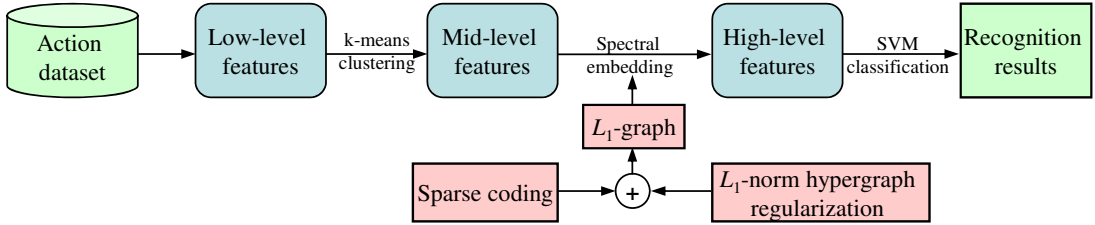

Fig. 1. The flowchart of human action recognition using the latent semantics (i.e. high-level features) learnt by spectral embedding based on the $L_1$-graph constructed with structured sparse representation.

level latent semantics from the abundant mid-level features. To our best knowledge, [19] is the first attempt to extract latent semantics from videos for human action recognition using a manifold learning technique based on diffusion maps [20]. Although this diffusion map method has been shown to achieve better results than the information theoretic models in [19], it requires fine parameter tuning for graph construction which can significantly affect the performance and has been noted as an inherent weakness of graph-based methods.

To address the above problems associated with human action recognition, we propose a novel latent semantic learning method based on spectral embedding [21]–[23] with $L_1$-graph, without the need to tune any parameter for graph construction as a key step of manifold learning. More importantly, we construct the $L_1$-graph with structured sparse representation, which can be obtained by structured sparse coding with its structured sparsity being ensured by novel $L_1$-norm hypergraph regularization over mid-level features. A distinct advantage of characterizing the similarity between mid-level features based on structured sparse representation is that we can collect a sparse affinity matrix in a parameter-free manner. In contrast, since the mid-level features are represented as vectors of point-wise mutual information and their similarity is typically characterized via a Gaussian function in [19], the choice of the variance in this function has been shown to affect the performance of human action recognition significantly. To summarize, through spectral embedding based on $L_1$-graph, we can discover more intrinsic manifold structure hidden among mid-level features and thus learn more compact but discriminative latent semantics by spectral clustering in the new embedding space, which has been shown in our later experiments. In this paper, we focus on parameter-free $L_1$-graph construction and only consider the commonly used spectral embedding method [21], regardless of many other manifold learning techniques in the literature.

Since our new $L_1$-norm hypergraph regularization can ensure the structured sparsity in $L_1$-graph construction, we discuss it in detail as follows. Although derived from the traditional Laplacian regularization [24]–[26], our $L_1$-norm hypergraph regularization is more suitable for parameter-free $L_1$-graph construction as an $L_1$-norm term (see Section III). More importantly, we can exploit the manifold structure of mid-level features for graph construction and simultaneously introduce another important type of sparsity by $L_1$-norm hypergraph regularization, which is the main difference between our structured sparse coding and the traditional sparse coding

[27]–[29]. In this paper, the hypergraph [30]–[32] used for our $L_1$-norm hypergraph regularization is also constructed in a parameter-free manner. That is, each video that contains multiple mid-level features is regarded as a hyperedge, and its weight can be estimated based on the original cluster centers associated with mid-level features. Although both spare representation and hypergraph are also used in our short conference version [29], the present paper has integrated them in a unified structured sparse representation framework. Since our $L_1$-norm hypergraph regularization can be applied to many other machine learning problems (considering the wide use of Laplacian regularization), the present paper has made a significant extra contribution as compared to [29]. In addition, the proposed new structured sparse coding can certainly be considered as another extra contribution.

In this paper, we apply the learnt latent semantics to human action recognition with SVM by defining a histogram intersection kernel. The flowchart of human action recognition is illustrated in Fig. 1, which contains four components: extraction of low-level spatio-temporal descriptors the same as [6], formation of mid-level features by $k$-means clustering, extraction of high-level latent semantics by spectral embedding based on $L_1$-graph, and action classification with SVM. We have tested our method on the commonly used KTH action dataset [5] and unconstrained YouTube action dataset [18]. The experimental results demonstrate the superior performance of our method for human action recognition. To emphasize the main contributions of this paper, we summarize the following advantages of our method:

(1) Our method can learn compact but discriminative latent semantics by exploring the manifold structure of visual keywords in both $L_1$-graph construction and spectral embedding, which is quite different from the traditional latent semantic analysis based on topic models.

(2) This is the first attempt to develop novel structured sparse coding for latent semantic learning in the challenging task of human action recognition, although many efforts have already been made to apply sparse coding to other difficult tasks in the literature.

(3) Our new $L_1$-norm hypergraph regularization can incorporate the manifold structure of mid-level features into graph construction. More importantly, it can be further applied to many other machine learning problems, considering the wide use of Laplacian regularization.

(4) Our method has been shown to significantly outperform other latent semantic learning approaches [13], [17]–[19], which turns to be more impressive given that we do not use feature pruning [7], [18], [19], multiple types of low-level features [7], [11], [18], or spatio-temporal layout information [11], [17], [33] for human action recognition.

The remainder of this paper is organized as follows. Section II gives a brief review of related work. Section III proposes a latent semantic learning method based on structured sparse representation. In Section IV, we present the details of human action recognition with SVM using our learnt latent semantics. In Section V, our method is evaluated on the commonly used KTH action dataset and unconstrained YouTube action dataset. Finally, Section VI gives the conclusions.

## II. RELATED WORK

Our method differs from other latent semantic learning approaches based on latent topic [13], [14] or information theoretic models [17], [18] in that the manifold structure of mid-level features can be explored in both $L_1$-graph construction and spectral embedding, which results in compact but discriminative high-level features. Although the diffusion map method [19] can also exploit this manifold structural information for latent semantic learning, it requires fine parameter tuning for graph construction which can significantly affect the performance. In contrast, our method can construct the $L_1$-graph in a parameter-free manner by structured sparse coding with $L_1$-norm hypergraph regularization. More importantly, as shown in later experiments, our spectral embedding with $L_1$-graph can help to discover more intrinsic manifold structure of mid-level features and thus learn more compact but discriminative latent semantics. Here, it should be noted that we focus on parameter-free graph construction for manifold learning in this paper and thus only adopt the commonly used spectral embedding method introduced in [21], without considering other manifold learning techniques developed in the literature.

Although our latent semantic learning method can be regarded as dimensionality reduction over mid-level features, it completely differs from the traditional dimensionality reduction approaches [20], [22] based on spectral embedding. Firstly, the latent semantics learnt by our method can help to form high-level representation and thus bridge the semantic gap to some extent. This is also the reason why the topic models [13], [14] for latent semantic analysis are widely used for multimedia information processing. However, the traditional dimensionality reduction approaches based on spectral embedding fail to give explicit explanation of each reduced feature by directly using the eigenvectors of the Laplacian matrix to form the new feature representation. Secondly, our latent semantic learning method by spectral embedding with $L_1$-graph over mid-level features incurs much less time cost than the traditional dimensionality reduction approaches by spectral embedding with graphs over all the data.

In the literature, many efforts have been made to explore sparse coding [27], [28] for different difficult tasks. However, this paper makes the first attempt to apply structured sparse coding to latent semantic learning for action recognition. More importantly, we have developed a novel structured sparse coding algorithm for $L_1$-graph construction with the structured sparsity being ensured by $L_1$-norm hypergraph regularization, different from the traditional $L_1$-graph construction methods [34], [35] without considering the structured sparsity. Here, it should be noted that our new $L_1$-norm hypergraph regularization is defined directly over all the eigenvectors of the hypergraph Laplacian matrix, other than the $p$-Laplacian regularization [36] as an ordinary $L_1$-generalization (with $p = 1$) of the Laplacian regularization. Moreover, although Laplacian regularization is also combined with sparse coding in visual keyword generation [37], it is just a quadratic term and thus is hard to be used in parameter-free $L_1$-graph construction for learning latent semantics from visual keywords. In contrast, our $L_1$-norm hypergraph regularization can be readily used for parameter-free $L_1$-graph construction as an $L_1$-norm term. We will provide further comparison to [36], [37] in Section III. In this paper, we focus on exploring the manifold structure of mid-level features in spare coding, regardless of other types of structured sparsity [38], [39].

Since our main goal is to learn compact but discriminative latent semantics from abundant mid-level features for human action recognition, we consider very simple experimental setting in this paper. For example, only a single type of low-level spatio-temporal descriptors are extracted from action videos just the same as [6]. Moreover, the learnt high-level features are directly applied to action recognition without considering their spatio-temporal layout information. In fact, we do not use feature pruning [7], [18], [19], multiple types of low-level features [7], [11], [18], or spatio-temporal layout information [11], [17], [33] for action recognition. However, even with such simple setting, our method can still achieve improvements with respect to the state of the arts, as shown in our later experiments.

## III. LATENT SEMANTIC LEARNING WITH STRUCTURED SPARSITY REPRESENTATION

In this section, we first propose a sparse coding algorithm to construct $L_1$-graph for spectral embedding over mid-level features. To explore structured sparsity in $L_1$-graph construction, we further improve the sparse coding algorithm with $L_1$-norm hypergraph regularization. Finally, in the new embedding space, we learn latent semantics from abundant mid-level features by spectral clustering.

### A. Spectral Embedding with $L_1$-Graph

Given a vocabulary of mid-level features $\mathcal{V}_m = \{m_i\}_{i=1}^M$, each video can be represented as a histogram of mid-level features $\{c_n(m_i) : i = 1, ..., M\}$, where $c_n(m_i)$ is the count of times that $m_i$ occurs in video $n$ ($n = 1, ..., N$). Based on this BOW representation, our goal is to discover the manifold structure of mid-level features by spectral embedding with graphs for learning compact but discriminative latent semantics, which is different from the topic models [13], [14] for latent semantic analysis. Although many spectral embedding methods have been developed in previous work, this paper focuses on graph construction as the key step of spectral embedding. That is, once a graph is constructed, we can adopt any spectral embedding method to discover the manifold structure hidden among mid-level features. Since the traditional graph construction method proposed in [19] has difficulty in choosing the variance for the Gaussian function, we thus construct a graph with sparse representation (i.e. $L_1$-graph) in a parameter-free manner, inspired by recent advances in sparse coding [27], [28]. Specifically, we first represent each mid-level feature $m_i$ $(i = 1, ..., M)$ as a vector $x_i = \{c_n(m_i) : n = 1, ..., N\}$ and then find the solution of linear reconstruction of $m_i$ using the rest of mid-level features based on sparse coding. Since the obtained sparse coefficients for linear reconstruction can be used to define the similarity between mid-level features, we succeed in constructing an $L_1$-graph for spectral embedding.

Our $L_1$-graph construction by linear reconstruction with sparse coding is presented in detail as follows. For each mid-level feature $m_i$ $(i = 1, ..., M)$, we suppose it can be reconstructed using the rest of mid-level features, which results in an underdetermined linear system: $x_i = B_i \alpha_i$, where $x_i \in R^N$ is the vector of $m_i$ to be approximated, $\alpha_i \in R^{M-1}$ is the vector for unknown reconstruction coefficients, and $B_i = [x_1, x_2, ..., x_{i-1}, x_{i+1}, ..., x_M] \in R^{N \times (M-1)}$ is the overcomplete dictionary with $M-1$ bases. According to [27], if the solution for $x_i$ is sparse enough, it can be recovered by:

$$\min_{\alpha_i} \ ||\alpha_i||_1, \ \text{s.t.} \ x_i = B_i \alpha_i, \tag{1}$$

where $||\alpha_i||_1$ is the $L_1$-norm of $\alpha_i$. Considering that the structured sparsity term (i.e. $L_1$-norm hypergraph regularization) is defined over all mid-level features, we need to use $x_i$ also as a base and thus reformulate the above spare representation problem as follows:

$$\min_{\alpha_i} \ ||C_i \alpha_i||_1, \ \text{s.t.} \ x_i = B \alpha_i, \tag{2}$$

where $\alpha_i \in R^M$ is the new vector for unknown reconstruction coefficients, $B = [x_1, ..., x_M] \in R^{N \times M}$ is the overcomplete dictionary with $M$ bases, and $C_i \in R^{M \times M}$ is a diagonal matrix with its $(j, j)$-element $C_i(j, j) = +\infty (j = i)$ and $C_i(j, j) = 1 (j \neq i)$. Due to such special form of $C_i$, we always have $\alpha_i(i) = 0$ for problem (2), where $\alpha_i(i)$ is the $i$-th element of $\alpha_i$. This means that we can obtain a solution equivalent to that of the original problem (1). Here, it should be noted that the distinct advantage of the above reformulation is that the $L_1$-norm hypergraph regularization defined over all mid-level features can now be readily explored in spare representation, which will be shown in the next subsection. Moreover, if we set $\tilde{\alpha}_i = C_i \alpha_i$, the spare representation problem (2) can be transformed into:

$$\min_{\tilde{\alpha}_i} \ ||\tilde{\alpha}_i||_1, \ \text{s.t.} \ x_i = BC_i^{-1} \tilde{\alpha}_i, \tag{3}$$

which takes the same form as the original problem (1). In practice, due to the noise, we can reconstruct $x_i$ similar to [28]: $x_i = BC_i^{-1} \tilde{\alpha}_i + \zeta_i$, where $\zeta_i$ is the noise term. The above problem can then be redefined by minimizing the $L_1$-norm of

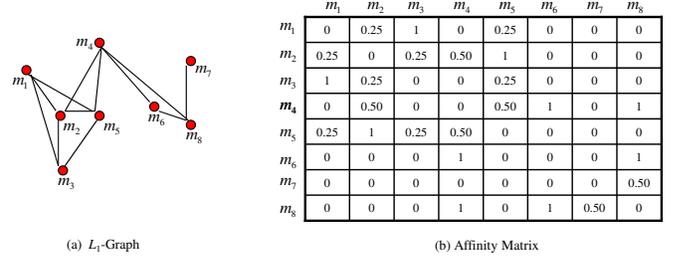

|       | $m_1$ | $m_2$ | $m_3$ | $m_4$ | $m_5$ | $m_6$ | $m_7$ | $m_8$ |
|-------|-------|-------|-------|-------|-------|-------|-------|-------|
| $m_1$ | 0     | 0.25  | 1     | 0     | 0.25  | 0     | 0     | 0     |
| $m_2$ | 0.25  | 0     | 0.25  | 0.50  | 1     | 0     | 0     | 0     |
| $m_3$ | 1     | 0.25  | 0     | 0     | 0.25  | 0     | 0     | 0     |
| $m_4$ | 0     | 0.50  | 0     | 0     | 0.50  | 1     | 0     | 1     |
| $m_5$ | 0.25  | 1     | 0.25  | 0.50  | 0     | 0     | 0     | 0     |
| $m_6$ | 0     | 0     | 0     | 1     | 0     | 0     | 0     | 1     |
| $m_7$ | 0     | 0     | 0     | 0     | 0     | 0     | 0     | 0.50  |
| $m_8$ | 0     | 0     | 0     | 1     | 0     | 1     | 0.50  | 0     |

(a) $L_1$-Graph　　(b) Affinity Matrix

Fig. 2. Illustration of the $L_1$-graph constructed by sparse coding. The vocabulary of mid-level features $\mathcal{V}_m = \{m_i\}_{i=1}^8$, and the set of five videos are represented as: video $1 = \{m_2, m_5\}$, video $2 = \{m_1, m_2, m_3\}$, video $3 = \{m_4, m_5, m_6\}$, video $4 = \{m_6, m_7, m_8\}$, and video 5 $= \{m_2, m_5, m_6, m_8\}$. If we assume that each mid-level feature occurs only once in a video, the affinity matrix given by Fig. 2(b) can be computed by sparse coding, and the corresponding $L_1$-graph is shown in Fig. 2(a) where each graph edge only denotes whether a pair of mid-level features are related and its length has no meaning.

both reconstruction coefficients and reconstruction error:

$$\min_{\alpha_i'} \ ||\alpha_i'||_1, \ \text{s.t.} \ x_i = B_i' \alpha_i', \tag{4}$$

where $B_i' = [BC_i^{-1}, I] \in R^{N \times (N+M)}$ and $\alpha_i' = [\tilde{\alpha}_i^T, \zeta_i^T]^T$. This convex optimization can be transformed into a general linear programming problem and thus has a globally optimal solution. After we have obtained the reconstruction coefficients for all the mid-level features, we can define an affinity matrix $A = \{a_{ij}\}_{M \times M}$ as follows (considering the special form of $C_i$):

$$a_{ij} = \begin{cases} |\alpha_i'(j)|, & j \neq i; \\ 0, & j = i, \end{cases} \tag{5}$$

where $\alpha_i'(j)$ is the $j$-th element of the vector $\alpha_i'$. By setting $A = (A + A^T)/2$, we can construct an undirected graph $\mathcal{G} = \{\mathcal{V}, A\}$ with the vertex set $\mathcal{V}$ using the vocabulary $\mathcal{V}_m$. In the following, we will called it as $L_1$-graph, since it is obtained by $L_1$-optimization.

Due to the sparse representation given by equation (4), each mid-level feature (i.e. vertex) in this $L_1$-graph is only related to several other mid-level features (see an example shown in Fig. 2). Although the traditional $k$-nearest neighbors ($k$-NN) graph also has such sparse property, our $L_1$-graph constructed by linear reconstruction with sparse coding has a distinct advantage, i.e., we can determine the number of related mid-level features automatically for each mid-level feature and thus do not need to set it as a fixed value like $k$-NN graph. For example, we can observe from Fig. 2(a) that each mid-level feature is related to different number of related mid-level features. Moreover, another advantage of our $L_1$-graph is that the similarity between mid-level featured is learnt by sparse coding in a parameter-free manner, while the traditional graph construction method proposed in [19] has difficulty in choosing the variance for the Gaussian function given that it is used to characterize the similarity measure.

Based on the above $L_1$-graph, we further perform spectral embedding to discover the manifold structure of mid-level features. The goal of spectral embedding is to represent each vertex in the $L_1$-graph as a lower dimensional vector that preserves the similarities between the vertex pairs. Actually, this is equivalent to finding the leading eigenvectors of the normalized Laplacian matrix

$$\mathcal{L} = I - D^{-1/2}AD^{-1/2}, \tag{6}$$

where $D$ is a diagonal matrix with its $(i,i)$-element equal to the sum of the $i$-th row of the affinity matrix $A$. In this paper, we only consider this type of normalized Laplacian [21], regardless of other normalized versions [20]. Let $\{(\lambda_i, \mathbf{v}_i) : i = 1, ..., M\}$ be the set of eigenvalues and the associated eigenvectors of $\mathcal{L}$, where $0 \leq \lambda_1 \leq ... \leq \lambda_M$ and $\mathbf{v}_i^T \mathbf{v}_i = 1$. The *spectral embedding* of the $L_1$-graph is given by

$$E = [\mathbf{v}_1, ..., \mathbf{v}_K], \tag{7}$$

where the $j$-th row $E_{j\cdot}$ of the matrix $E$ can be regarded as the new representation for vertex $m_j$. Here, it should be noted that we focus on parameter-free $L_1$-graph construction for manifold learning in this paper and thus only adopt the spectral embedding method introduced in [21], without considering other manifold learning techniques that have been developed in the literature. Since we usually set $K < M$, the mid-level features have actually been represented as lower dimensional vectors which can be further used for latent semantic learning by spectral clustering.

### B. $L_1$-Graph Construction with Structured Sparsity Representation

In the above $L_1$ graph, the similarity between mid-level features is defined as the reconstruction coefficients of the linear reconstruction solution obtained by sparse coding. However, the structured sparsity of these reconstruction coefficients is ignored in such sparse representation. In this paper, we only consider one special type of structure, i.e., the manifold structure of the mid-level features. Actually, this manifold structure can be explored in sparse representation based on the normalized Laplacian matrix of the hypergraph [30]–[32] defined over mid-level features, which is well known as Laplacian regularization or hypergraph regularization. The distinct advantage of hypergraph regularization is that the structured sparsity can be ensured for sparse representation and thus we can obtain new structured sparse representation for $L_1$-graph construction. Since the hypergraph plays an important role in structured sparse representation, we will first give the details of hypergraph construction.

In fact, the hypergraph can be constructed in a parameter-free manner. That is, each video that contains multiple mid-level features is regarded as a hyperedge, and its weight can be estimated based on the original cluster centers associated with mid-level features. Suppose each video is represented as a histogram of mid-level features $\{c_j(m_i) : i = 1, ..., M\}$, where $c_j(m_i)$ is the count of times that mid-level feature $m_i$ occurs in video $j$ ($j = 1, ..., N$). The hypergraph $\mathcal{G} = \{\mathcal{V}, \mathcal{E}, \mathbf{w}\}$ can be constructed as follows. We first set $\mathcal{V} = \mathcal{V}_m = \{m_i\}_{i=1}^M$ and $\mathcal{E} = \{e_j : e_j = \{m_i : c_j(m_i) > 0, i = 1, ..., M\}\}_{j=1}^N$. The incidence matrix $H$ of the hypergraph $\mathcal{G}$ can be directly defined by

$$H_{ij} = c_j(m_i) / \sum_{m_{i'} \in e_j} c_j(m_{i'}). \tag{8}$$

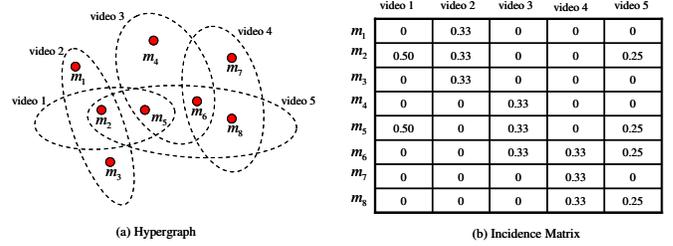

Fig. 3. Illustration of the hypergraph constructed in a parameter-free manner. In Fig. 3(a), each dashed ellipse denotes a hyperedge (i.e. video), and each red solid node denotes the vertex (i.e. mid-level feature). The incidence matrix $H$ of the hypergraph given by Fig. 3(b) is computed using the occurrences of mid-level features within videos.

Here, we consider a soft incidence matrix (i.e. $H_{ij} \in [0, 1]$), which is different from [31] with $H_{ij} = 1$ or 0. Moreover, we define the hyperedge weights $\mathbf{w} = \{w(e_j)\}_{j=1}^N$ by

$$w(e_j) = \frac{1}{|e_j|} \sum_{m_i \in e_j, m_{i'} \in e_j} R_{ii'}, \tag{9}$$

where $|e_j|$ denotes the number of vertices within $e_j$, and $R$ is the linear kernel matrix defined with the original cluster centers associated with mid-level features. This ensures that the weight of $e_j$ is set to a larger value when this hyperedge is more compact. Given these hyperedge weights, we can define the degree of a vertex $m_i \in \mathcal{V}$ as $d(m_i) = \sum_{e_j \in \mathcal{E}} w(e_j) H_{ij}$. For a hyperedge $e_j \in \mathcal{E}$, its degree is defined as $\delta(e_j) = \sum_{m_i \in \mathcal{V}} H_{ij}$. An example hypergraph is shown in Fig. 3.

It is worth noting that the above hypergraph construction method is parameter-free, which is similar to our $L_1$-graph construction method by linear reconstruction with sparse coding. More importantly, according to [30], the above hypergraph can capture the high order correlation between mid-level features. Moreover, to define the hypergraph regularization term for sparse representation, we first compute the normalized Laplacian matrix the same as [31]:

$$\mathcal{L}_h = I - D_v^{-1/2} H W D_e^{-1} H^T D_v^{-1/2}, \tag{10}$$

where $Dv$, $De$, and $W$ denote the diagonal matrices of the vertex degrees, the hyperedge degrees, and the hyperedge weights, respectively. Based on this normalized Laplacian matrix $\mathcal{L}_h$, we can then define the hypergraph regularization term for the sparse representation problem (2) as $\alpha_i^T \mathcal{L}_h \alpha_i$, which can also be regarded as a smoothness measure of $\alpha_i$ over the hypergraph.

However, this hypergraph regularization term is hard to be directly incorporated into the sparse representation problem (2), no matter as a part of the objective function or a constraint condition. Hence, we further formulate an $L_1$-norm version of hypergraph regularization as:

$$||B_h \alpha_i||_1 = ||\Sigma_h^{\frac{1}{2}} V_h^T \alpha_i||_1, \tag{11}$$

where $B_h = \Sigma_h^{\frac{1}{2}} V_h^T$, $V_h$ is an $M \times M$ orthonormal matrix with each column being an eigenvector of $\mathcal{L}_h$, and $\Sigma_h$ is an $M \times M$ diagonal matrix with its diagonal element $\Sigma_h(i, i)$ being an eigenvalue of $\mathcal{L}_h$ (sorted as $\Sigma_h(1, 1) \leq ... \leq \Sigma_h(M, M)$). Given that $\mathcal{L}_h$ is nonnegative definite, $\Sigma_h \geq 0$ (i.e. all the eigenvalues $\geq 0$). Since $\mathcal{L}_h V_h = V_h \Sigma_h$ and $V_h$ is orthonormal, we have $\mathcal{L}_h = V_h \Sigma_h V_h^T$. Hence, the original hypergraph regularization can be reformulated as:

$$\alpha_i^T \mathcal{L}_h \alpha_i = \alpha_i^T V_h \Sigma_h^{\frac{1}{2}} \Sigma_h^{\frac{1}{2}} V_h^T \alpha_i = \alpha_i^T B_h^T B_h \alpha_i = ||B_h \alpha_i||_2^2, \quad (12)$$

which means that our new formulation $||B_h \alpha_i||_1$ is indeed an $L_1$-norm version of the original hypergraph regularization. By introducing noise terms for linear reconstruction and $L_1$-norm hypergraph regularization, we transform the sparse representation problem (2) into

$$\min_{\alpha_i, \zeta_i, \xi_i} \quad ||[(C_i \alpha_i)^T, \zeta_i^T, \xi_i^T]||_1,$$
$$\text{s.t.} \quad x_i = B\alpha_i + \zeta_i, \quad 0 = B_h \alpha_i + \xi_i, \quad (13)$$

where the reconstruction error and hypergraph smoothness with respect to $\alpha_i$ are controlled by $\zeta_i$ and $\xi_i$, respectively. If we set $\tilde{\alpha}_i = C_i \alpha_i$, we can reformulate the above problem as

$$\min_{\tilde{\alpha}_i, \zeta_i, \xi_i} \quad ||[\tilde{\alpha}_i^T, \zeta_i^T, \xi_i^T]||_1,$$
$$\text{s.t.} \quad x_i = BC_i^{-1}\tilde{\alpha}_i + \zeta_i, \quad 0 = B_h C_i^{-1}\tilde{\alpha}_i + \xi_i, \quad (14)$$

Let $\alpha_i' = [\tilde{\alpha}_i^T, \zeta_i^T, \xi_i^T]^T$, $B_i' = \begin{bmatrix} BC_i^{-1} & I & 0 \\ B_h C_i^{-1} & 0 & I \end{bmatrix}$, and $x_i' = [x_i^T, 0^T]^T$. We finally solve the following structured spare representation problem for $L_1$-graph construction:

$$\min_{\alpha_i'} \quad ||\alpha_i'||_1, \quad \text{s.t.} \quad x_i' = B_i'\alpha_i', \quad (15)$$

which takes the same form as the original spare representation problem (4). The affinity matrix $A$ of the $L_1$-graph can be defined the same as equation (5).

In our new formulation of structured sparse representation, our $L_1$-norm hypergraph regularization can be smoothly incorporated into the original sparse representation problem (2). However, this is not true for the tradition hypergraph regularization or Laplacian regularization, which may introduce extra parameters into the $L_1$-optimization and thus has conflict with our original goal of parameter-free $L_1$-graph construction. Moreover, our $L_1$-norm hypergraph regularization can cause another type of sparsity (see the extra noise term $\xi_i$), which can not be ensured by the tradition Laplacian regularization. These are also the main differences between our structured spare coding for high-level latent semantic learning and the Laplacian sparse coding proposed in [37] for mid-level feature generation. Here, it should be noted that the $p$-Laplacian regularization [36] can also be regarded as an ordinary $L_1$-generalization of the Laplacian regularization with $p = 1$. By defining a matrix $C_p \in R^{\frac{M(M-1)}{2} \times M}$, the $p$-Laplacian regularization can be formulated as $||C_p \alpha_i||_1$ [40], similar to our $L_1$-norm hypergraph regularization. Hence, we can apply $p$-Laplacian regularization similarly to structured sparse representation. However, it incurs too large time cost due to the large matrix $C_p$ even for a moderate vocabulary size $M = 500$.

### C. Latent Semantic Learning by Spectral Clustering

After the $L_1$-graph has been constructed with structured spare representation, we perform spectral embedding using the normalized Laplacian matrix. In the new low-dimensional embedding space, we learn high-level latent semantics by spectral clustering. The algorithm is summarized as follows:

(1) Find $K$ smallest nontrivial eigenvectors $\mathbf{v}_1, ..., \mathbf{v}_K$ of the normalized Laplacian matrix $\mathcal{L}$ of the $L_1$-graph constructed with structured spare representation.

(2) Form $E = [\mathbf{v}_1, ..., \mathbf{v}_K]$, and normalize each row of $E$ to have unit length. Here, the $i$-th row $E_i$ is a new low-dimensional feature vector for mid-level feature $m_i$.

(3) Perform $k$-means clustering on the new feature vectors $E_i, (i = 1, ..., M)$ to partition the vocabulary $\mathcal{V}_m$ of $M$ mid-level features into $K$ clusters. Here, each cluster of mid-level features denotes a new high-level feature.

In the following, our latent semantic learning algorithm based on spectral embedding with structured sparse representation will be denoted as S²LSL (i.e. structured sparse latent semantic learning), while the algorithm based on spectral embedding only with sparse representation will be denoted as SLSL (i.e. sparse latent semantic learning). Since the spectral embedding is performed with $L_1$-graph over mid-level features, our algorithm can run efficiently even on a large video dataset.

## IV. HUMAN ACTION RECOGNITION WITH SVM

In this section, we present the details of human action recognition with SVM using our learnt latent semantics. We first derive a new semantics-aware representation (i.e. histogram of high-level features) for each video from the original BOW representation, and then define a histogram intersection kernel based on the new representation for action cognition with SVM.

Let $\mathcal{V}_h = \{h_i\}_{i=1}^{K}$ be the vocabulary of high-level features learnt from the vocabulary of mid-level features $\mathcal{V}_m = \{m_j\}_{j=1}^{M}$ by our S²LSL or SLSL algorithm. The BOW representation with $\mathcal{V}_h$ for each video can be derived from the original BOW representation with $\mathcal{V}_m$ as follows. Given the count of times $c_n(m_j)$ that mid-level feature $m_j$ occurs in video $n$, the count of times $c_n(h_i)$ that high-level feature $h_i$ occurs in this video can be computed by:

$$c_n(h_i) = \sum_{j=1}^{M} c_n(m_j)c(m_j, h_i), \quad (16)$$

where $c(m_j, h_i) = 1$ if mid-level feature $m_j$ occurs in cluster $i$ (i.e. high-level feature $h_i$) according to the above spectral clustering and $c(m_j, h_i) = 0$ otherwise. That is, each video is now represented as a histogram of high-level features. Similar to the traditional BOW representation, the above semantics-aware representation can be used to define a histogram intersection kernel $K_{HI}$:

$$K_{HI}(n, \tilde{n}) = \sum_{i=1}^{K} \min(c_n(h_i), c_{\tilde{n}}(h_i)). \quad (17)$$

This semantics-aware kernel $K_{HI}$ is further used for human action recognition with SVM.

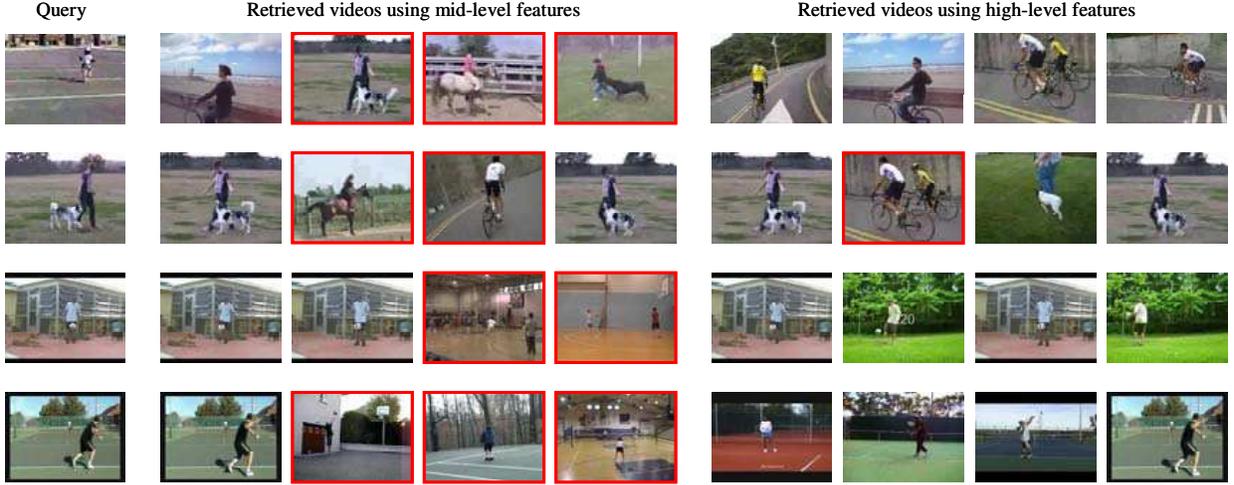

Fig. 4. Retrieval examples using mid-level and high-level features on the YouTube action dataset [18]. For each query, four videos with the highest values of the histogram intersection kernel are retrieved. The incorrectly retrieved videos (which do not come from the same action category as the query) are marked with red boxes. The high-level features are shown to achieve significantly better retrieval results than the mid-level features.

To provide preliminary evaluation of our learnt latent semantics, we apply the above semantics-aware kernel to action retrieval, and some retrieval examples on the YouTube action dataset [18] are shown in Fig. 4. Here, we learn 400 high-level features from 2,000 mid-level features by our S²LSL algorithm. We can find that the high-level features can achieve significantly better retrieval results than the mid-level features, which means that the learnt high-level features can provide a semantically more succinct representation but a more discriminative descriptor of human actions than the mid-level features. Moreover, we can also find in the experiments that similar dominating high-level features are used to represent the videos from the same action category, although their exact meanings are unknown. This is also the reason why we call them "latent semantics" in this paper like the traditional topic models. In the following, we will apply our semantics-aware representation to human action recognition on the commonly used KTH action dataset [5] and unconstrained YouTube action dataset [18].

## V. EXPERIMENTAL RESULTS

In this section, our latent semantic learning method will be evaluated on two standard action datasets. We first describe the experimental setup, including information of the two action datasets and the implementation details. Moreover, we compare our latent semantic learning method with other closely related methods on the two standard action datasets, respectively.

### A. Experimental Setup

We select two different action datasets for performance evaluation. The first dataset is KTH [5] which contains six actions: boxing, clapping, waving, jogging, running, and walking. These actions are performed by 25 actors under four different scenarios. In total, this dataset contains 598 video clips. Since KTH has been widely used for performance evaluation in human action recognition, we can make direct comparison with the state-of-the-art methods using their own results on this dataset. The second dataset is YouTube [18] which has lots of camera movement, cluttered backgrounds, different viewing directions, and varying illumination conditions. Hence, it is significantly more complex and challenging than KTH. This action dataset contains 11 categories: diving, golf swinging (g_swinging), horse riding (h_riding), soccer juggling (s_juggling), swinging, tennis swinging (t_swinging), trampoline jumping (t_jumping), volleyball spiking (v_spiking), basketball shooting (b_shooting), biking, and walking (with a dog). Most of them share some common motions such as "jumping" and "swinging". The video clips are organized into 25 relatively independent groups, where separate groups are either taken in different environments or by different photographers. The dataset contains 1,168 video clips in total. To the best of our knowledge, this is one of the most extensive realistic action datasets in the literature.

To extract low-level features from the two action datasets, we adopt the spatio-temporal interest point detector proposed in [6]. Compared to the 3D Harris-Corner detector [8], it generates dense features which can improve the recognition performance in most cases. Specifically, this detector makes use of 2D Gaussian filter and 1D Gabor filters in spatial and temporal directions, respectively. A response value is given at every position $(x, y, t)$. The interest points are selected at the locations of local maximal responses, and 3D cuboids are extracted around them. For simplicity, we describe the 3D cuboids using the flat gradient vectors, which are further reduced to 100 dimensions by PCA the same as [18], [19]. In our experiments, we extract 400 descriptors from each video clip for the KTH dataset, while for the YouTube dataset more descriptors (i.e. 1,600) are extracted from each vide clip since this dataset is more complex and challenging. Finally, on the two action datasets, we quantize the extracted spatio-temporal descriptors into $M$ mid-level features by $k$-means clustering. Here, it should be noted that we adopt very simple experimental setting for low-level feature extraction, since we focus on learning compact but discriminative latent semantics in this paper. We do not consider pruning low-level features [7], [18], [19] or combining multiple types of low-level features [7], [11], [18] for human action recognition. However, even with such simple setting, our latent semantic learning method can still achieve performance improvements with respect to the state of the arts, as shown in our later experiments.

Since the diffusion map (DM) method for latent sematic learning proposed in [19] has been reported to outperform other manifold learning techniques (e.g. Isomap [41] and Eigenmaps [42]) and also the information theoretic approaches (e.g. information maximization [17]), we focus on comparing our $S^2LSL$ with DM and do not make direct comparison with [17], [41], [42]. In fact, our $S^2LSL$ has been shown in later experiments to perform much better than DM, and thus we succeed in verifying the superiority of our method indirectly with respect to other manifold learning techniques and the information theoretic approaches. Moreover, to show the effectiveness of our structured sparse representation, we also compare our $S^2LSL$ with SLSL that does not consider the structured sparsity. Finally, our $S^2LSL$ is compared with LDA and BOW, since they are the most widely used in the literature. Here, all the methods for comparison except BOW are designed to learn latent semantics from a large vocabulary of mid-level features. In the following, we select $M = 2,000$ for the four latent semantic learning methods (i.e. $S^2LSL$, SLSL, DM, and LDA). For the two action datasets, we use 24 actors or groups for training SVM and the rest for testing, just the same as previous work [18], [19].

### B. Results on the KTH Dataset

The comparison of the four latent semantic learning methods is shown in Fig. 5. We find that our $S^2LSL$ generally performs the best. As compared to SLSL without considering the structured sparsity, our $S^2LSL$ leads to better results in most cases, which means that the structured sparsity ensured by our new $L_1$-norm hypergraph regularization is very useful for learning compact but discriminative latent semantics. In fact, the better discriminative ability of the high-level features learnt by our $S^2LSL$ may be due to the fact that the manifold structure of mid-level features can be explored by $L_1$-norm hypergraph regularization in $L_1$-graph construction and then latent semantic learning. Moreover, we also find that our $S^2LSL$ can always achieve performance improvement over the DM method for latent semantic learning [19], which becomes more significant when the number of high-level features is relatively smaller (e.g. $K \leq 150$). The reason may be that our $S^2LSL$ has eliminated the need to tune any parameter for graph construction based on structured sparse representation, while DM heavily suffers from the difficulty of parameter tuning in graph construction since the Gaussian function is used to characterize the similarity between mid-level features. Here, it should be noted that such parameter tuning can significantly affect the performance and has been noted as an inherent weakness of graph-based methods.

Similar to topic models such as LDA, our $S^2LSL$ can explicitly learn latent semantics from abundant mid-level features.

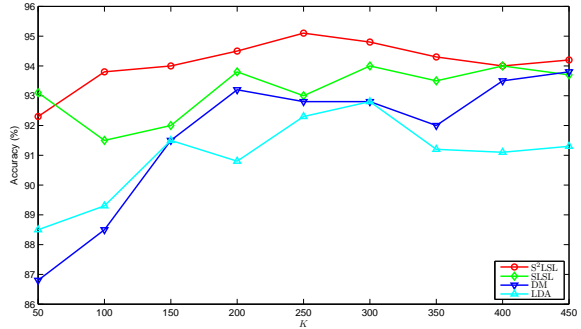

Fig. 5. Comparison of the four latent semantic learning methods for human action recognition on the KTH dataset.

TABLE I
THE RELATIVE PERFORMANCE OF OUR $S^2LSL$ AS COMPARED TO BOW ($M = 2,000$) ON THE KTH DATASET

| $K/M$ (%) | 2.5 | 5.0 | 7.5 | 10.0 | 12.5 | 15.0 | 17.5 | 20.0 | 22.5 |
|---|---|---|---|---|---|---|---|---|---|
| Speed Gain (%) | 440 | 337 | 282 | 243 | 215 | 191 | 172 | 155 | 143 |
| Accuracy Gain (%) | -1.0 | 0.6 | 0.9 | 1.4 | 2.0 | 1.7 | 1.2 | 0.9 | 1.1 |

However, since the manifold structure of mid-level features can be explored in both $L_1$-graph construction and spectral embedding, our $S^2LSL$ is able to generate more compact but discriminative high-level features for human action recognition, which can be observed from Fig. 5. Specifically, our $S^2LSL$ is shown to outperform LDA significantly in all cases, which could be due to that our learnt high-level features have better discriminative ability. Moreover, we can also find that our $S^2LSL$ consistently achieves promising results with varied number of high-level features, while LDA suffers from obvious performance degradation for a large number of high-level features since the model parameters of LDA increase as $K$ grows and thus only local optima may be found in this case.

In the above experiments, we learn latent semantics from $M = 2,000$ mid-level features. To demonstrate the gain achieved by our $S^2LSL$, we need to make direct comparison to BOW with $M = 2,000$. Table I shows the relative performance of our $S^2LSL$, where both speed and accuracy gains are computed relatively upon BOW. In particular, to obtain the speed gain, we compare the speed of kernel computation on the high-level features learnt by our $S^2LSL$ to that of kernel computation on the 2,000 mid-level features. Here, we only consider kernel computation since the SVM classification incurs the same time cost once the kernel matrix is provided. From Table I, we can observe that our $S^2LSL$ can reduce the number of features to a very low level (e.g. 5.0%) without obvious performance degradation (or even with performance improvement), which is exactly consistent with the original goal of latent semantic learning in the literature. This nice property of our $S^2LSL$ can speed up the subsequent classification and retrieval significantly, which is extremely important for large datasets.

Our $S^2LSL$ is further compared to BOW with $M = 2,000$ on each action category. Here, we only consider $K = 250$ for our $S^2LSL$. To make extensive comparison, we take BOW with $M = 250$ as a baseline method. The comparison between our S²LSL and these two BOW methods is shown in Fig. 6. We can find that our S²LSL leads to improvements over BOW ($M = 2,000$) on four action categories: "boxing", "waving", "jogging", and "running", without performance degradation on the other categories, even when the number of features is decreased from 2,000 to 250. The ability of our HSE to achieve promising results using only a small number of features is important because it means that the proposed method is scalable for large datasets. Moreover, our S²LSL is shown to perform better than BOW ($M = 300$) on all the action categories when they select the same number of features.

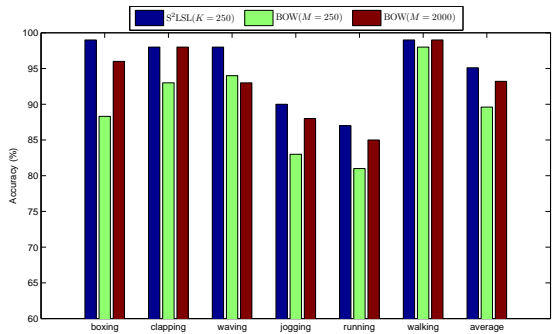

Fig. 6. Comparison between S²LSL and BOW for human action recognition on the KTH dataset. Here, our S²LSL learns 250 high-level features (i.e. $K = 250$) from 2,000 mid-level feature (i.e. $M = 2,000$).

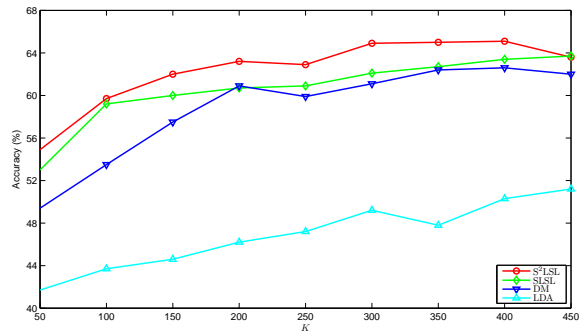

Fig. 7. Comparison of the four latent semantic learning methods for human action recognition on the YouTube dataset.

| Methods | FP | MF | STL | Accuracy (%) |
|---|---|---|---|---|
| Dollar et al. [6] | no | no | no | 81.2 |
| Niebles et al. [13] | no | no | no | 83.3 |
| Liu and Shah. [17] | no | no | yes | 94.2 |
| Bregonzio et al. [7] | yes | yes | no | 93.2 |
| Liu et al. [18] | yes | yes | no | 93.8 |
| Liu et al. [19] | yes | no | no | 92.3 |
| Oikonomopoulos et al. [33] | no | no | yes | 88.0 |
| Wu et al. [11] | no | yes | yes | 94.5 |
| Our method | no | no | no | 95.1 |

| $K/M$ (%) | 2.5 | 5.0 | 7.5 | 10.0 | 12.5 | 15.0 | 17.5 | 20.0 | 22.5 |
|---|---|---|---|---|---|---|---|---|---|
| Speed Gain (%) | 973 | 701 | 544 | 441 | 365 | 308 | 268 | 235 | 207 |
| Accuracy Gain (%) | -14.1 | -6.6 | -3.0 | -1.1 | -1.5 | 1.6 | 1.7 | 1.9 | -0.5 |

Since we focus on learning compact but discriminative latent semantics for human action recognition, we consider very simple experimental setting in this paper. For example, in the experiments, only a single type of low-level spatio-temporal descriptors are extracted from action videos the same as [6]. Moreover, the learnt high-level features are directly applied to human action recognition without considering their spatio-temporal layout information. That is, we do not use feature pruning [7], [18], [19], multiple types of low-level features [7], [11], [18], or spatio-temporal layout information [11], [17], [33] for action recognition. However, even with such simple experimental setting, our S²LSL method can still achieve performance improvements with respect to the state of the arts, as shown in Table II. This also provides further convincing validation of the effectiveness of our latent semantic learning method based on structured sparse representation by $L_1$-norm hypergraph regularization.

### C. Results on the YouTube Dataset

The YouTube dataset is more complex and challenging than KTH, since it has lots of camera movement, cluttered backgrounds, different viewing directions, and varying illumination conditions. We repeat the same experiments on this dataset, and the recognition results are shown in Fig. 7, Table III, and Fig. 8. Here, the four latent semantic learning methods are compared in Fig. 7, while in Table III and Fig. 8 we focus on comparing our S²LSL directly with BOW to show the relative gain achieved by our S²LSL. The speed gain in Table III is still computed when only kernel computation is concerned. Overall, we can make the same observations on this dataset as we have done with the KTH dataset.

Specifically, our S²LSL can generally achieve better performance than the other latent semantic learning approaches. This observation further verifies that our S²LSL can learn more compact but discriminative latent semantics by exploring the manifold structure of mid-level features in both graph construction and spectral embedding. Moreover, the compact set of high-level features learnt by our S²LSL can speed up the subsequent kernel computation significantly without obvious performance degradation (when $K/M > 5.0\%$), as shown in Table III. In particular, when our S²LSL ($K = 400$) is compared to BOW ($M = 400$ or $2,000$) on each action category, we can observe from Fig. 8 that our S²LSL leads to performance improvements over BOW on most action categories. The reason may be that the high-level features learnt by our S²LSL can help to reduce the semantic ambiguity of the most confusing action categories. When we focus on the comparison between S²LSL ($K = 400$) and BOW ($M = 2,000$), the performance improvement achieved by our S²LSL is really impressive given that we have decreased the number of features from 2,000 to 400. Although the commonly used LDA can do the same thing as our our S²LSL, it completely fails in this case, as shown in Fig. 7. Considering the superior performance of LDA reported in the literature and also the promising results reported in this paper, our S$^2$LSL can be regarded as the best among different representative latent semantic learning approaches.

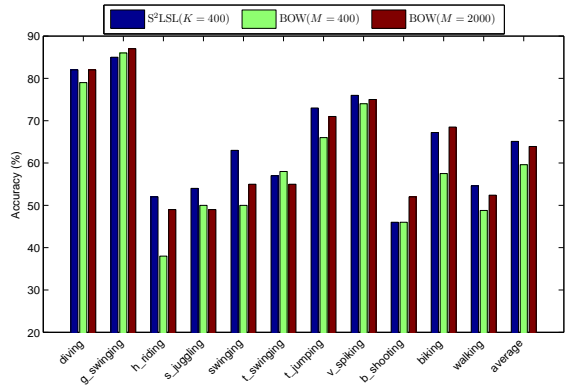

Fig. 8. Comparison between S$^2$LSL and BOW for human action recognition on the YouTube dataset. Here, our S$^2$LSL learns 400 high-level features (i.e. $K = 400$) from 2,000 mid-level feature (i.e. $M = 2,000$).

## VI. CONCLUSIONS

We have investigated the challenging problem of latent semantic learning in the application of human action recognition. To bridge the semantic gap associated with human action recognition, we have proposed a novel latent semantic learning method based on structured sparse representation. To exploit the manifold structure of mid-level features for latent semantic learning, we have developed a spectral embedding approach based on the $L_1$-graph constructed with structured sparse representation in a parameter-free manner, without the need to tune any parameter for graph construction. Although many efforts have been made to explore both sparse representation and hypergraph in different applications in the literature, we have made the first attempt to integrate them in a unified structured sparse representation framework for latent semantic learning. The experimental results have demonstrated the superior performance of our latent semantic learning method. In the future work, considering the wide use of Laplacian regularization, our new $L_1$-norm hypergraph regularization will be explored in many other machine learning problems to take into the structured sparsity into account. Moreover, our latent semantic learning method will be extended to other difficult tasks such as image annotation and scene classification.

## ACKNOWLEDGEMENTS

The work described in this paper was fully supported by the National Natural Science Foundation of China under Grant Nos. 60873154 and 61073084.

## REFERENCES

[1] P. Turaga, R. Chellappa, V. Subrahmanian, and O. Udrea, "Machine recognition of human activities: A survey," *IEEE Trans. Circuits and Systems for Video Technology*, vol. 18, no. 11, pp. 1473–1488, Nov. 2008.

[2] W.-L. Lu, K. Okuma, and J. Little, "Tracking and recognizing actions of multiple hockey players using the boosted particle filter," *Image and Vision Computing*, vol. 27, no. 1-2, pp. 189–205, 2009.

[3] D. Ramanan and D. Forsyth, "Automatic annotation of everyday movements," in *Advances in neural information processing systems 16*, 2004, pp. 1547–1554.

[4] V. Parameswaran and R. Chellappa, "View invariance for human action recognition," *International Journal of Computer Vision*, vol. 66, no. 1, pp. 83–101, Jan. 2006.

[5] C. Schuldt, I. Laptev, and B. Caputo, "Recognizing human actions: A local SVM approach," in *Proc. ICPR*, vol. 3, 2004, pp. 32–36.

[6] P. Dollar, V. Rabaud, G. Cottrell, and S. Belongie, "Behavior recognition via sparse spatio-temporal features," in *Proc. 2nd Joint IEEE International Workshop on Visual Surveillance and Performance Evaluation of Tracking and Surveillance*, 2005, pp. 65–72.

[7] M. Bregonzio, S. Gong, and T. Xiang, "Recognising action as clouds of space-time interest points," in *Proc. CVPR*, 2009, pp. 1948–1955.

[8] I. Laptev, M. Marszalek, C. Schmid, and B. Rozenfeld, "Learning realistic human actions from movies," in *Proc. CVPR*, 2008.

[9] L. Ballan, M. Bertini, A. D. Bimbo, L. Seidenari, and G. Serra, "Effective codebooks for human action categorization," in *Proc. ICCV International Workshop on Video-Oriented Object and Event Classification*, 2009, pp. 506–513.

[10] A. Kovashka and K. Grauman, "Learning a hierarchy of discriminative space-time neighborhood features for human action recognition," in *Proc. CVPR*, 2010, pp. 2046–2053.

[11] X. Wu, D. Xu, L. Duan, and J. Luo, "Action recognition using context and appearance distribution features," in *Proc. CVPR*, 2011, pp. 489–496.

[12] Y. Mu, J. Sun, T. Han, L.-F. Cheong, and S. Yan, "Randomized locality sensitive vocabularies for bag-of-features model," in *Proc. ECCV*, 2010, pp. 748–761.

[13] J. Niebles, H. Wang, and L. Fei-Fei, "Unsupervised learning of human action categories using spatial-temporal words," *International Journal of Computer Vision*, vol. 79, no. 3, pp. 299–318, Sept. 2008.

[14] Y. Wang and G. Mori, "Human action recognition by semilatent topic models," *IEEE Trans. Pattern Analysis and Machine Intelligence*, vol. 31, no. 10, pp. 1762–1774, Oct. 2009.

[15] T. Hofmann, "Unsupervised learning by probabilistic latent semantic analysis," *Machine Learning*, vol. 41, pp. 177–196, 2001.

[16] D. Blei, A. Ng, and M. Jordan, "Latent Dirichlet allocation," *Journal of Machine Learning Research*, vol. 3, pp. 993–1022, 2003.

[17] J. Liu and M. Shah, "Learning human actions via information maximization," in *Proc. CVPR*, 2008.

[18] J. Liu, J. Luo, and M. Shah, "Recognizing realistic actions from videos in the wild," in *Proc. CVPR*, 2009, pp. 1996–2003.

[19] J. Liu, Y. Yang, and M. Shah, "Learning semantic visual vocabularies using diffusion distance," in *Proc. CVPR*, 2009, pp. 461–468.

[20] S. Lafon and A. Lee, "Diffusion maps and coarse-graining: A unified framework for dimensionality reduction, graph partitioning, and data set parameterization," *IEEE Trans. Pattern Analysis and Machine Intelligence*, vol. 28, no. 9, pp. 1393–1403, 2006.

[21] A. Ng, M. Jordan, and Y. Weiss, "On spectral clustering: Analysis and an algorithm," in *Advances in Neural Information Processing Systems 14*, 2002, pp. 849–856.

[22] S. Yan, D. Xu, B. Zhang, H. Zhang, Q. Yang, and S. Lin, "Graph embedding and extensions: A general framework for dimensionality reduction," *IEEE Trans. Pattern Analysis and Machine Intelligence*, vol. 29, no. 1, pp. 40–51, 2007.

[23] H. Wang, S. Yan, T. Huang, and X. Tang, "Maximum unfolded embedding: formulation, solution, and application for image clustering," in *Proc. ACM Multimedia*, 2006, pp. 45–48.

[24] X. Zhu, Z. Ghahramani, and J. Lafferty, "Semi-supervised learning using Gaussian fields and harmonic functions," in *Proc. ICML*, 2003, pp. 912–919.

[25] D. Zhou, O. Bousquet, T. Lal, J. Weston, and B. Schölkopf, "Learning with local and global consistency," in *Advances in Neural Information Processing Systems 16*, 2004, pp. 321–328.

[26] R. Ando and T. Zhang, "Learning on graph with Laplacian regularization," in *Advances in Neural Information Processing Systems 19*, 2007, pp. 25–32.

[27] D. Donoho, "For most large underdetermined systems of linear equations the minimal $\ell^1$-norm solution is also the sparsest solution," *Communications on Pure and Applied Mathematics*, vol. 59, no. 7, pp. 797–829, 2004.

[28] J. Wright, A. Yang, A. Ganesh, S. Sastry, and Y. Ma, "Robust face recognition via sparse representation," *IEEE Trans. Pattern Analysis and Machine Intelligence*, vol. 31, no. 2, pp. 210–227, 2009.

[29] Z. Lu, Y. Peng, and H. Ip, "Spectral learning of latent semantics of action recognition," in *Proc. ICCV*, 2011.

[30] S. Agarwal, K. Branson, and S. Belongie, "Higher order learning with graphs," in *Proc. ICML*, 2006, pp. 17–24.

[31] D. Zhou, J. Huang, and B. Schölkopf, "Learning with hypergraphs: Clustering, classification, and embedding," in *Advances in Neural Information Processing Systems 19*, 2007, pp. 1601–1608.

[32] L. Sun, S. Ji, and J. Ye, "Hypergraph spectral learning for multi-label classification," in *Proc. KDD*, 2008, pp. 668–676.

[33] A. Oikonomopoulos, I. Patras, and M. Pantic, "Spatiotemporal localization and categorization of human actions in unsegmented image sequences," *IEEE Trans. Image Processing*, vol. 20, no. 4, pp. 1126–1140, Apr. 2011.

[34] B. Cheng, J. Yang, S. Yan, and T. Huang, "Learning with $\ell^1$-graph for image analysis," *IEEE Trans. Image Processing*, vol. 19, no. 4, pp. 858–866, Apr. 2010.

[35] H. Cheng, Z. Liu, and J. Yang, "Sparsity induced similarity measure for label propagation," in *Proc. ICCV*, 2009, pp. 317–324.

[36] D. Zhou and B. Schölkopf, "Regularization on discrete spaces," in *Proc. DAGM*, 2005, pp. 361–368.

[37] S. Gao, I. Tsang, L.-T. Chia, and P. Zhao, "Local features are not lonely - Laplacian sparse coding for image classification," in *Proc. CVPR*, 2010, pp. 3555–3561.

[38] E. Elhamifar and R. Vidal, "Robust classification using structured sparse representation," in *Proc. CVPR*, 2011, pp. 1873–1879.

[39] R. Jenatton, J. Mairal, G. Obozinski, and F. Bach, "Proximal methods for hierarchical sparse coding," *Journal of Machine Learning Research*, vol. 12, pp. 2297–2334, 2011.

[40] X. Chen, Q. Lin, S. Kim, J. Carbonell, and E. Xing, "An efficient proximal gradient method for general structured sparse learning," 2010. [Online]. Available: http://arxiv.org/abs/1005.4717

[41] M. Balasubramanian and E. Schwartz, "The isomap algorithm and topological stability," *Science*, vol. 295, no. 5552, pp. 7–7, 2002.

[42] M. Belkin and P. Niyogi, "Laplacian eigenmaps for dimensionality reduction and data representation," *Neural Computation*, vol. 15, no. 6, pp. 1373–1396, 2003.

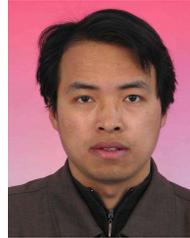

**Zhiwu Lu** received the M.Sc. degree in applied mathematics from Peking University, Beijing, China in 2005, and the Ph.D. degree in computer science from City University of Hong Kong in 2011.

Since March 2011, he has become an assistant professor with the Institute of Computer Science and Technology, Peking University. He has published over 30 papers in international journals and conference proceedings including TIP, TSMC-B, TMM, AAAI, ICCV, CVPR, ECCV, and ACM-MM. His research interests lie in machine learning, computer vision, and multimedia information retrieval.

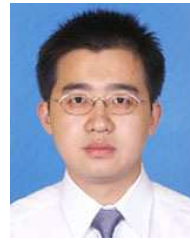

**Yuxin Peng** received the Ph.D. degree in computer science and technology from Peking University, Beijing, China, in 2003.

He joined the Institute of Computer Science and Technology, Peking University, as an assistant professor in 2003 and was promoted to a professor in 2010. From 2003 to 2004, he was a visiting scholar with the Department of Computer Science, City University of Hong Kong. His current research interests include content-based video retrieval, image processing, and pattern recognition.